\documentclass[floatfix,showpacs,amsmath,amssymb,letterpaper,groupaddresses,superscriptaddress]{article}
\usepackage{graphicx}
\usepackage{latexsym}
\usepackage{amsmath}
\usepackage{amssymb}
\usepackage{epsfig}
\usepackage{graphicx}
\usepackage{graphicx}
\usepackage{amssymb,amsmath,times}
\usepackage{hyperref}
\usepackage{color}

\def\be{\begin{equation}}
\def\ee{\end{equation}}
\def\ba{\begin{eqnarray}}
\def\ea{\end{eqnarray}}
\def\la{\langle}
\def\ra{\rangle}

\usepackage{epsfig}
\usepackage{graphicx}
\usepackage{multirow}
\usepackage{caption}
\begin{document}
\begin{center}
{\Large \bf On the robustness of topological quantum codes:\\ Ising perturbation\\
}
\vspace{1cm}
Mohammad Hossein Zarei\footnote{email:mzarei92@shirazu.ac.ir}\\
\vspace{5mm}
\vspace{1cm} Physics Department, College of Sciences, Shiraz University, Shiraz 71454, Iran\\
\end{center}
\vskip 3cm
\begin{abstract}
We study the phase transition from two different topological phases to the ferromagnetic phase by focusing on points of the phase transition. To this end, we present a detailed mapping from such models to the Ising model in a transverse field. Such a mapping is derived by re-writing the initial Hamiltonian in a new basis so that the final model in such a basis has a well-known approximated phase transition point. Specifically, we consider the toric codes and the color codes on some various lattices with Ising perturbation. Our results provide a useful table to compare the robustness of the topological codes and to explicitly show that the robustness of the topological codes depends on triangulation of their underlying lattices.
\end{abstract}
\section{Introduction}
One of the important problems in the control of quantum systems is the decoherence which emerges from undesirable interactions \cite{decoherence}. Specially in quantum computation processes, it is very important to find a quantum memory which is robust against emergent errors. In recent years, finding a solution for the decoherence problem has been one of the biggest challenges \cite{dec2, dec3}. One of the options to circumvent this problem is topological quantum computation where topological nature of a physical system is used to perform a fault tolerant quantum computation \cite{kitaev, kit2}.\\
The key property in the topological quantum computational model is topological order. Topological order is a new phase of matter which for the first time was presented by Wen in 1980 \cite{wen}. It is completely different from ordinary order such as ferromagnetic order so that the topological orders cannot be described by Landau paradigm and symmetry breaking theory \cite{land}. Also there is no local order parameter which recognizes topological phase. In fact topological order is described by a non-local order parameter and it is independent from local details of physical systems \cite{wen2}. There are many physical systems in condensed matter physics with topological order \cite{cond}, for example in quantum Hall effects \cite{hall}, high-temperature superconductors \cite{sup, sup2, sup3} and frustrated
magnetism \cite{mag, mag2, mag3}. \\
Among physical systems with topological order, lattice models have been taken into more consideration, because of their simplicity. In 1997, Kitaev presented a model Hamiltonian on a torus which could encode the subspace of two qubits in it's ground states \cite{kitaev}. Thereafter, topological color codes were introduced by Bombin et al. \cite{cc, cc1, cc2, sqoc} where the unitary Clifford gates could be topologically applied while this was not possible in the toric code.\\ Degeneracy of the ground state in the topological codes depend on topological nature of the model and local small perturbations cannot lift such a degeneracy. The robustness of the degeneracy of the ground state is only guaranteed in small perturbations so that when perturbation increases, at a critical value, a phase transition occurs. In fact, there is a topological phase transition (TPT) where the point of phase transition is a measure of the robustness of the code. The robustness of the topological codes are studied against thermal perturbations and local perturbations \cite{1, 2, 3, 4, 5, 6}.  As an example, robustness of the toric codes against parallel and transverse magnetic fields have been studied where magnetic field can be treated as an extrinsic factor which perturbs topological order \cite{m2, m3, m4, m5}. If the perturbation is as Ising interaction, there is a phase transition between topological phase and the ferromagnetic phase. Such a problem has been studied for the toric code in an Ising perturbation (TCI) on a square lattice \cite{karimipour} and for the color code in Ising perturbation (CCI) on the honeycomb lattice in \cite{kargar}.\\
In this paper, we focus on the effect of the Ising perturbation on topological codes on various lattices. We show that such models are mapped to Ising model in transverse field (IT) by re-writing the Hamiltonian in a new basis. Since the approximated phase transition points for the IT on different lattices are well-known \cite{Ising, Ising2, Ising3}, we provide a useful table to compare the phase transition points of topological codes on various lattices.\\We consider dependence of the phase transition point and also the robustness of topological codes on triangulation of their underlying lattices. In fact, for a pure topological model on a specific lattice, there are some topological properties which are independent of triangulation of the lattice \cite{kitaev}. When the perturbation is added, the system is not topological. However, before the phase transition occurs, topological properties are still independent of triangulation. Since the point of phase transition is a property of the perturbed topological code, it is expected that the phase transition point depends on triangulation. We derive points of phase transition for perturbed topological codes on various triangulations and explicitly show dependence of phase transition points on triangulation of the lattice.\\
In the first step, we compare the robustness of CCI's on two specific lattices. We show that they are mapped to Ising model in transverse field on two different lattices with different phase transition points. The same result is derived for the TCI's on three different lattices. We emphasize that although it is well known that the phase transition point in an ordinary quantum phase transition depends on the underlying lattice, such a problem has not been explicitly studied for a TPT.\\
In the second step, we compare the robustness of the TCI and CCI. Such a work has been performed for the topological codes in a parallel field. In \cite{4}, the authors have compared the robustness of the color code and the toric code in parallel field. They have concluded that the color code is more robust than the toric code against parallel magnetic field. We show that underlying lattices of these codes are important in such a problem so that the robustness of TCI on square lattice is less than the robustness of the CCI on honeycomb lattice while the TCI on honeycomb lattice is as robust as the CCI on honeycomb lattice. So the robustness of topological codes is very sensitive to triangulation of the lattices and we cannot compare different topological codes without considering the underlying lattices. The structure of this paper is as follows:\\
In Section (\ref{sec1}) we review topological properties of two topological codes, the color codes and the toric codes. In Section (\ref{sec2}) we apply Ising perturbation to the topological color codes on various lattices. Then we apply the same perturbation to the toric code. In this section we present a detailed mapping to re-write Hamiltonian of the models in a new basis so that the initial models are converted to the Ising model in the transverse field. Finally, in Section (\ref{sec3}), we provide a table of the phase transition points and compare the robustness of the topological codes.
\section{Topological code states}\label{sec1}
In this section, we review some important properties of the topological quantum codes. Specifically, we consider two well-known topological codes, the color codes and the toric codes. Topological order of these models is a non-local property which does not depend on the underlying lattices. For example the color codes can be defined on the honeycomb and square-octagonal lattices so that, in both cases, topological order is the same.
\subsection{Topological color codes }
\begin{figure}[t]
\centering
\includegraphics[width=4cm,height=3.5cm,angle=0]{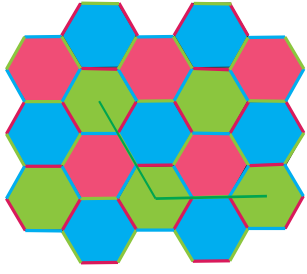}
\caption{(Color online) In the color code on the honeycomb lattice, spins live on vertices of the lattice. The lattice is colored by three different colors, red, blue and green. Each link connects two plaquettes with the same color and it is colored by the same color, so all links are colored by three colors, red, blue and green. A green string move on green links and it has two end points in two green plaquettes.} \label{hexagonal}
\end{figure}

\begin{figure}[t]
\centering
\includegraphics[width=5cm,height=4.5cm,angle=0]{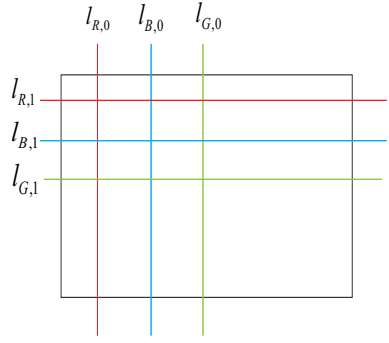}
\caption{ (Color online) Schematically we show three colored loops which wind around torus in two different directions.
} \label{loops}
\end{figure}

In this subsection, we review the topological color codes (TCC). It has been shown that they can be used for topological implementation of Clifford unitary gates and it is an advantage of these codes compared with the toric codes \cite{cc1}. \\
The TCC models are defined on three-colorable lattices where one can color all faces (plaquettes) of the lattice by three different colors so that any two neighbor plaquettes are not in the same color \cite{cc2}. Here we consider the TCC model on a honeycomb lattice, but we emphasize that topological properties are the same for all TCC models. To this end, consider a honeycomb lattice on a torus where spins live on vertices of the lattice, see Figure (\ref{hexagonal}). There are two kinds of operators corresponding to each plaquette of the lattice as follows:
\begin{equation}
P_{x}=\prod_{i \in P}X_{i}~~~,~~~P_{z}=\prod_{i \in P}Z_{i},
\end{equation}
where $X$ and $Z$ denote the Pauli operators $\sigma _x $ and $\sigma _z$, respectively. If the number of the plaquettes is $N$, the number of the plaquette operators will be equal to $2N$. The most important problem is to find the stabilized subspace of the code which is defined as follows:
\begin{equation}
\mathcal{L}=\{ |\psi\ra~~ |~~ P_x |\psi\ra =|\psi\ra , P_z |\psi\ra =|\psi\ra \}.
\end{equation}
This subspace can also be considered as the subspace of ground states of a Hamiltonian as:
\begin{equation}\label{top}
H=-\sum_{p} P_x - \sum_{p}P_z .
\end{equation}
By the fact that $P_x (1+P_x )=(1+P_x)$, it is simple to check that the following state is a ground state of the Hamiltonian (\ref{top}):
\begin{equation}\label{00}
|\psi\ra =\prod_{p} (1+P_{x})|00...0\ra ,
\end{equation}
where $|0\ra$ is eigenstate of the Pauli operator $Z$. Also there are many constraints between the plaquette operators. As it has been shown in figure (\ref{hexagonal}), the honeycomb lattice is colored by three colors such as blue, green, red. According to the periodic boundry conditions on the torus, it is clear that the following constraints will be hold on the plaquette operators:
\begin{equation}\label{dd}
\prod_{P\in R} P_{x}=\prod_{P\in G}P_{x}=\prod_{P\in B}P_{x}~~,~~\prod_{P\in R} P_{z}=\prod_{P\in G}P_{z}=\prod_{P\in B}P_{z},
\end{equation}
where $P\in R$ refers to the plaquttes which are colored by red and so on. Hence, from $2N$ plaquette operators, only $2N-4$ number of them are independent. Because the number of qubits is also equal to $2N$, degeneracy of the ground state is sixteen-fold. \\In the following, we show that all ground states are constructed by applying some non-local operators to state (\ref{00}). For defining these non-local operators, we consider that all edges of the honeycomb lattice can also be colored by three colors so that each edge with a specific color connects two plaquettes with the same color, see figure(\ref{hexagonal}). Then we define three kinds of color strings on the lattice where each string moves on the edges and plaquettes with the same color and two endpoints of the strings live in the plaquettes, see figure (\ref{hexagonal}). We denote these strings by $S^{R},S^{G},S^{B}$. We define two string operators $S_z ^{C} , S_x ^{C}$ corresponding to each string in the form of:
\begin{equation}
S_z ^{C} = \prod_{i\in S^{C}}Z_i~~,~~S_x ^{C} = \prod_{i\in S^{C}}X_i ,
\end{equation}
where the Pauli operators are applied on all qubits which live on the string $S^{C}$ and $C$ can be each one of three colors, red, green or blue. If two end points of a string are connected to each other, it generates a closed loop and we will have a corresponding loop operator. Since each plaquette shares in two spins with the loop operator, it is simple to show each loop operator commutes with all plaquette operators in the Hamiltonian (\ref{top}). By attention to topology of the torus, the loop operators are divided to two general classes. One class contains trivial loops which are contractible and another contains non-trivial loops which are non-contractible.\\
Important point is about non-trivial operators. There are six fundamental non-trivial loops which are denoted by three colors and two directions on the torus, see figure (\ref{loops}). Related to each kind of the non-trivial loops, there are two operators which are constructed by $Z$ or $X$ operators. We denote these operators as:
\begin{equation}
L^{C,\sigma}_{x}=\prod_{i\in l_{C,\sigma}}X_i ~~,~~L^{C,\sigma}_{z}=\prod_{i\in l_{C,\sigma}}Z_i ,
\end{equation}
where $l_{C,\sigma}$ refers to a non-trivial loop and $C$ denotes color of the loop and $\sigma=\{0,1\}$ denotes direction of the loop around the torus. An important property of the non-trivial loop operators is that they cannot be written as product of the plaquette operators in the Hamiltonian (\ref{top}). But it is simple to show from three loop operators $L^{R,\sigma}, L^{G,\sigma}$ and $L^{B,\sigma}$, only two number of them are independent and one can be written as product of two other ones and some plaquette operators. So we select only two colors of these loops, for example red and green.\\
Because of the topology of the torus, two non-trivial loops where one is in direction $0$ and another is in direction $1$ cross each other, see Figure (\ref{loops}), so there is an anti-commutation relation between the non-trivial loop operators as follows:
\begin{equation}\label{q}
\{L^{R,0}_x ,L^{B,1}_z \}=0~~,~~\{L^{R,1}_x ,L^{B,0}_z \}=0~~,~~\{L^{B,0}_x ,L^{R,1}_z \}=0~~,~~\{L^{B,1}_x ,L^{R,0}_z \}=0 .
\end{equation}
Other relations between these operators are commutation. According to the above relations the non-trivial loop operators generate a $2^4$ dimensional space which can encode the space of four qubits. Also we can explicitly construct all sixteen bases of this space by the non-local operators in relation (\ref{q}). In fact, it is clear that the state $|\psi\ra$ as (\ref{00}) is stabilized by $L^{B,1}_z$, $L^{B,0}_z$, $L^{R,1}_z$, $L^{R,0}_z$, because they commute with operators $P_x$ then they are applied to state $|00...0\ra$ which is eigenstate of the Pauli operator $Z$. So all bases of the degenerate subspace are constructed as follows:
\begin{equation}
|\psi_{ijkl}\ra=(L^{B,1}_x)^i .(L^{B,0}_x)^j .(L^{R,1}_x)^k .(L^{R,0}_x)^l |\psi \ra ,
\end{equation}
where $i,j,k$ and $l$ are 0 or 1. We should emphasize that degeneracy of the ground state in this model is related to the topology of torus and the colorable structure of the lattice.
\subsection{The toric codes}
The most simple of the topological codes are the toric codes which can be defined on each oriented graph \cite{karim}. Here we consider a honeycomb lattice on a torus with $N$ spins which live on the edges of the lattice. The toric code on such a lattice is defined by the following plaquette and vertex operators:
\begin{equation}
B_p = \prod_{i\in p} Z_i ~~~,~~~A_v =\prod_{i\in v} X_i ,
\end{equation}
\begin{figure}[t]
\centering
\includegraphics[width=4cm,height=3.5cm,angle=0]{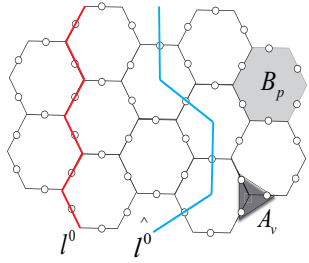}
\caption{(Color online) In the toric code on the honeycomb lattice, spins live on edge of the lattice, vertex operators are applied on three spins which are neighbors of each vertex and plaquette operators are applied on six spins which are on the edges of each plaquette. Two non-trivial loops on lattice has been shown, one moves on edges of lattice which has been denoted by red string and another moves on edges of dual lattice which has been denoted by blue string.
} \label{Kitaev}
\end{figure}
where $i \in p$ refers to spins which belong to plaquette $p$ and $i \in v$ refers to spins which are neighbor of vertex $v$, see Figure (\ref{Kitaev}). The stabilized subspace of this code can treated as the ground state of a Hamiltonian as:
\begin{equation}
H_{K}=-\sum_{p} B_p -\sum_{v} A_v .
\end{equation}
Since $A_v (1+A_v )= 1+A_v $, it is simple to show that the following un-normalized state is a ground state of this Hamiltonian:
\begin{equation}\label{as}
|\phi\ra =\prod_{v}(1+A_{v})|00...0\rangle .
\end{equation}
Also there are two constraints between the plaquette and vertex operators in the Hamiltonian as $\prod_{v} A_v =I$ and $\prod_{p}B_p =I$. So the number of independent operators is equal to $N-2$ and the ground state has four-fold degeneracy. To find other ground states we should consider two non-trivial loops $l^0 , l^1 $ around two directions of the torus where loops can be defined on the edges of lattice as well as the edges of dual lattice where the vertices, edges and faces of dual lattice are in one to one correspondence with the faces, edges and vertices of the lattice, respectively. There are two operators corresponding to a non-trivial loop $l^{\sigma}$ on the lattice and a non-trivial loop $\hat{l}^{\sigma}$ on the dual lattice, see figure (\ref{Kitaev}), in the form of:
\begin{equation}
L^{\sigma, x}=\prod_{i\in \hat{l}^{\sigma}}X_i ~~~,~~~L^{\sigma, z}=\prod_{i\in l^{\sigma}}Z_i .
\end{equation}
On one hand these operators commute with the Hamiltonian so they have joint eigenstates, and furthermore there exist the following anti-commutation relations between these operators:
\begin{equation}
\{L^{0, x},L^{1, z}\}=0~~,~~\{L^{0, z},L^{1, x}\}=0 .
\end{equation}
Other relations between these operators are commutation. According to the above relations all four ground states of the model are derived as:
\begin{equation}
(L_{0, x})^{i} .(L_{1, x})^{j}|\phi \ra ,
\end{equation}
where $i,j$ are 0 or 1, and $|\phi\ra$ is a state as in Eq.(\ref{as}). Thus in the toric codes, topological order of the ground states is associated to the periodic conditions of the torus and it is independent from the underlying lattice.\\
\subsection{The robustness of the topological codes}
The degenerate ground state of the topological codes can be used as a robust memory for quantum computation goals. For example, consider a perturbation as magnetic field on spins in the toric code. Such a perturbation leads to splitting between the degenerate states. The important point in this case is that two ground states of the model are only mapped to each other by a non-local operator as product of Pauli operators on a non-trivial loop which is related to the topology of the model. Therefore, if the perturbation is small enough, such a splitting cannot happen. \\
It is clear that the robustness of the degeneracy is lifted when perturbation increases. In fact a perturbation even in the first order leads to a transition term between the ground state and the exited states in the perturbed energy levels. This problem leads to a phase transition from topological order to an ordinary order. Finally, the topological order is lifted at a phase transition point. Therefore, the point of phase transition is a measure of the robustness of the topological codes.\\
\section{Ising perturbation on the topological code states }\label{sec2}
In this section, we consider the topological codes on different two-dimensional lattices with Ising perturbation. We show that these models are mapped to the Ising model in transverse field by re-writing the initial Hamiltonian in a non-local basis. In a general form, the Hamiltonian of the topological code with Ising perturbation is in the form of:
\begin{equation}\label{to}
H=-Jh_{Top}-kh_I
\end{equation}
where $h_{Top}$ is the Hamiltonian of the topological code and $h_I$ is the Hamiltonian of the Ising perturbation. At the first, let us consider two special regimes of the couplings. In case that $J=0$ or $K=\infty$, we have the Ising model which have a two-fold degenerate ground state where $Z$ components of all spins are up or down. In this case system is in the ferromagnetic phase and a local order parameter can describe phase of the model \cite{land}. Now we consider a case that $J=\infty$ or $K=0$, so we have the topological model where the system is in a topological phase and any local order parameter cannot describe such a phase \cite{wen}. Since in these two different regimes we have two different phases, it is a reasonable expectation that at a critical ratio of $k/J$ a phase transition occurs.\\
We re-write the Hamiltonian (\ref{to}) in a new basis and we show that in this new basis such a Hamiltonian is mapped to the Ising model in the transverse field on a 2D lattice.\\
\subsection{The topological color code on the honeycomb lattice}
\begin{figure}[t]
\centering
\includegraphics[width=8cm,height=3.5cm,angle=0]{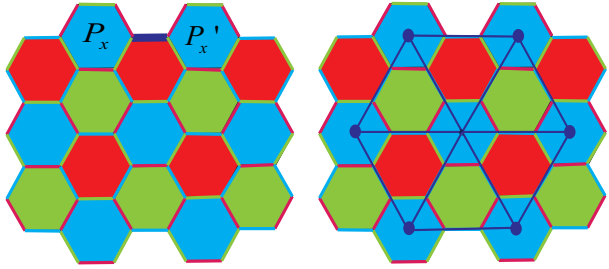}
\caption{(Color online) Left: An Ising perturbation corresponding to a blue link does not commute with two plaqutte operators $P_x$ and $P'_x$ in two endpoints of that link. Right: The virtual spins which live in blue plaquettes are denoted by blue circles. An Ising interaction $Z_i Z_j $ between two real spins on blue links does not commute with two plaquette operators $P_x$ on two it's end points and it is equivalent to an Ising interaction $X_p X_{p'}$ between virtual spins which live on corresponding plaquettes. Finally we have an Ising model on a triangular lattice.
} \label{triangular}
\end{figure}

Our method in studying the topological codes in presence of the Ising perturbation is to re-write Hamiltonian of the model in a new basis. It leads to convert the initial model to an Ising model in the transverse field. Similar method has been already used in other papers about the toric code on square lattice \cite{karimipour} and the color code on the honeycomb lattice \cite{kargar}. In this subsection, we explain the main idea for the TCC model on the honeycomb lattice then we apply it to other models on various lattices. To this end, we consider a Hamiltonian as:
\begin{equation}\label{topI}
H=-J(\sum_{p} P_x +\sum_{p}P_z)-K\sum_{\la i,j \ra} Z_i Z_j ,
\end{equation}
where $\la i,j \ra$ denotes neighbor spins on the honeycomb lattice. In this model, we have considered a positiv coupling constant $J$ for the TCC Hamiltonian and a positiv coupling constant $K$ for the Ising perturbation.\\
At the first, by the fact that the Ising perturbation commutes with the operators $P_z$ in the TCC model, we find that ground state of the new model is in a subspace which is stabilized by all operators $P_z$. We re-write the Hamiltonian (\ref{topI}) in such a subspace by choosing an un-normalized basis as follows:
\begin{equation}\label{qqq}
|\phi_{r_1 , r_2 , r_3 ,..., r_N}\ra=\prod_{P} (1+(-1)^{r_p}P_x)|00...0\rangle .
\end{equation}
where $N$ is the number of plaquettes of the lattice and $r_p$'s are binary numbers which are attributed to each plaquette. In fact we can attribute a virtual spin to each plaquette where $r_p$ denotes the value of the virtual spin in a plaquette $p$. The states in Eq(\ref{qqq}) are non-local so that they correspond to anyonic excitations of the TCC model. \\The number of states as (\ref{qqq}) is $2^{N}$ and all of them are stabilized by the operators $P_z$. About the above basis, we should emphasize that all these states are not independent, in fact there are constraints between the plaquette operators in different colors according to relation (\ref{dd}). For example product of all operators $P_{x}^{R}$ and $P_{x}^{G}$ is equal to the Identity and we find that $P_{x}^{R}.P_{x}^{G}|\phi_{r_1 , r_2 , r_3 ,..., r_N}\ra=|\phi_{r_1 , r_2 , r_3 ,..., r_N}\ra $. This basis would be complete if we add some new states which are constructed by applying the non-trivial loop operators on the state $|\phi\ra$.\\
In the following we should re-write the Hamiltonian (\ref{topI}) in the basis (\ref{qqq}). As we explained, all states (\ref{qqq}) are stabilized by $P_z$'s, so $P_z|\phi_{r_1 , r_2 , r_3 ,..., r_N}\ra=|\phi_{r_1 , r_2 , r_3 ,..., r_N}\ra$ and since the number of $P_z$'s is equal to N, we find that $\sum_{p}P_z |\phi_{r_1 , r_2 , r_3 ,..., r_N}\ra=N|\phi_{r_1 , r_2 , r_3 ,..., r_N}\ra$.\\
Then we apply the operator $P_x$ on the states (\ref{qqq}). By the fact that $P_x (1+(-1)^{r_p}P_x)=(-1)^{r_{p}}(1+(-1)^{r_p}P_x)$, we find that $P_x |\phi_{r_1 , r_2 , r_3 ,..., r_N}\ra=(-1)^{r_p}|\phi_{r_1 , r_2 , r_3 ,..., r_N}\ra$. We can describe this final result in an interesting view so that effect of $P_x$ in the basis (\ref{qqq}) is equivalent to effect of the Pauli operator $Z_p$ in a basis as $|r_1 , r_2 , ..., r_N\ra$ on the virtual spins. \\
Interesting situation occurs when we apply the Ising term on the basis (\ref{qqq}). To this end, as we explained in the previous section, we can color all links of the honeycomb lattice in three colors. Thus we divide the Ising interactions to three parts corresponding to the color of their links as:
\begin{equation}
H_I = \sum_{\la i,j\ra \in B} Z_i Z_j +\sum_{\la i,j\ra \in G} Z_i Z_j +\sum_{\la i,j\ra \in R} Z_i Z_j ,
\end{equation}
where $\la i,j \ra \in B $ denotes sites $i,j$ which are in two endpoints of a blue link and so on. For example, we apply the Ising interaction on the blue links to the state $|\phi_{r_1 , r_2 , r_3 ,..., r_N}\ra$. Each link $\la i,j\ra$ has four neighbor plaquettes and it is simple to show that corresponding operator to $\la i,j \ra$ as $Z_i Z_j$ does not commute with two plaquette operators $P_x$ corresponding to two plaquettes in two endpoints of that link, see Figure(\ref{triangular}, left). Therefore, we find that $Z_i Z_j (1+(-1)^{r_p}P_x)=(1+(-1)^{r_p +1}P_x)Z_i Z_j$, where $P_x$ is corresponding operator to one of the end points of the edge $\la i,j \ra$ . Finally, we conclude when the Ising interaction corresponding to a blue link of the lattice is applied to basis $|\phi_{r_1 , r_2 , r_3 ,..., r_N}\ra$, it rises up the  virtual spins $r_p , r_{p'}$ corresponding to two blue plaquettes in two end points of that link. Therefore, we can describe the Ising interaction $Z_i Z_j$ as an Ising interaction $X_pX_{p'}$ on the basis $|r_1 , r_2 , ..., r_N\ra$, where $p,p'$ are two blue plaquettes which are connected by a blue link. As you can see in Figure (\ref{triangular}, right), if we denote this Ising interaction between the neighbor virtual spins by a link, we will have a triangular lattice where all virtual spins in blue plaquettes live on the vertices of that lattice. \\
The above argument can also be repeated for the Ising interactions on the green and red links where, corresponding to each color, we will have an Ising model on a triangular lattice. \\
By considering all terms in the Hamiltonian (\ref{topI}) in the new basis, we will have a new Hamiltonian in the form of:
\begin{equation}\label{tran}
H=-JN-\sum_{\{p\}\in B}(JZ_p+KX_p X_{p'})-\sum_{\{p\}\in R}(JZ_p+KX_p X_{p'})-\sum_{\{p\}\in G}(JZ_p+KX_p X_{p'}) ,
\end{equation}
where $\{P\} \in B$ refers to all virtual spins on the blue plaquettes and $P ,P'$ refer to the neighbor plaquettes in the same color on a triangular lattice. For finding the ground state of the Hamiltonian, we use this fact that the final Hamiltonian in (\ref{tran}) on the virtual spins has been written as summation of three independent Hamiltonian on three triangular lattices with different colors. Therefore, it is enough to find the ground state of each part, independently. Interesting result is that each part of the Hamiltonian is an Ising model in a transverse field on a triangular lattice which a quantum phase transition has been known for it at $J/K \approx 4.77$ \cite{Ising}. At $J \ggg K$ limit all virtual spins are in positive direction of $Z$ and correspondingly the real spins are in ground state of the color code. At $K \ggg J$ limit All virtual spins are in the ground state of the Ising model on the triangular lattice and correspondingly the real spins are in the ground state of the Ising model on the honeycomb lattice.
\\
\subsection{the color code on the square-octagonal lattice}
\begin{figure}[t]
\centering
\includegraphics[width=4cm,height=3.5cm,angle=0]{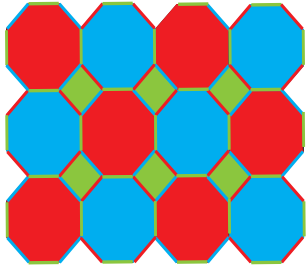}
\caption{(Color online) In the color code on Square-octagonal lattice, spins live on vertices of the lattice. Similar to the honeycomb lattice, we color all plaquettes and links by three different colors, red, blue and green.
} \label{sqhec}
\end{figure}

\begin{figure}[t]
\centering
\includegraphics[width=4cm,height=3.5cm,angle=0]{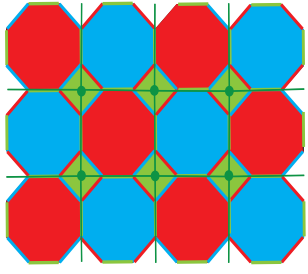}
\caption{(Color online) The virtual spins which live in green plaquettes are denoted by green circles. An Ising interaction $Z_i Z_j $ between two real spins on green links does not commute with two plaquette operators $P_x$ on two it's end points and it is equivalent to an Ising interaction $X_p X_{p'}$ between virtual spins which live on corresponding plaquettes. Finally we have an Ising model on a green square lattice.
} \label{sq}
\end{figure}

\begin{figure}[t]
\centering
\includegraphics[width=4cm,height=3.5cm,angle=0]{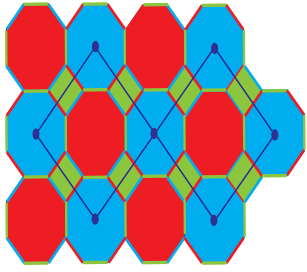}
\caption{ (Color online) The virtual spins which live in blue plaquettes are denoted by blue circles. There are two Ising interactions $Z_i Z_j $ between real spins on two blue links which connect two blue plaquettes. By the fact that these two Ising interactions does not commute with two plaquette operators $P_x$ on two it's end points, we find that they are equivalent to an Ising interaction $X_p X_{p'}$ between virtual spins which live on corresponding plaquettes. Finally we have an Ising model on a square lattice.
} \label{sq2}
\end{figure}

The color codes can be defined on any three-colorable lattices. Definition of the stabilizers are the same for all of them so that there are two kinds of stabilizers corresponding to each plaquette $P_x , P_z$ which are as product of $X$ or $Z$ operators on the spins of each plaquette. \\In this subsection, we consider the color code on a square-octagonal lattice in a Ising perturbation, see Figure (\ref{sqhec}). In order to find the phase transition point, similar to the previous section, we re-write the Hamiltonian of the model in the basis (\ref{qqq}). All of arguments are similar to the previous section so that here we divide the Hamiltonian to three parts corresponding to three colors of the plaquettes of the lattice. Each part of the Hamiltonian in the basis (\ref{qqq}) is as an Ising model in the transverse field on the virtual spins in plaquettes with the same color. Only distinction is kind of the Ising interaction pattern on the virtual spins in each color.\\
 As it has been shown in Figure (\ref{sq}), links of the lattice with green color generate a square lattice where the virtual spins live on it's vertices. Moreover, blue links generate a square lattice where the virtual spins live on it's vertices, see Figure(\ref{sq2}). Also we should be care that there are two blue links between each two blue neighbor plaquettes. This final point leads to an important distinction so that two Ising interactions in original model on the real spins correspond to an Ising interaction in the square lattice on the virtual spins. By considering a similar argument for the red plaquettes, the final Hamiltonian on the virtual spins will be as follows:
\begin{equation}
H=-KN- \sum_{\{p\} \in G} (JZ_p + K X_p X_{p'})- \sum_{\{p\} \in R}( JZ_p + 2K X_p X_{p'})- \sum_{\{p\} \in B} (JZ_p + 2K X_p X_{p'}),
\end{equation}
where $\{P\} \in G$ refers to all virtual spins on the green plaquettes and $P ,P'$ are neighbor plaquettes in the same color which generate a square lattice. Each part of the above Hamiltonian is an Ising model in the transverse field on a square lattice. For the first part, phase transition occurs at $J/K \approx  3$ \cite{Ising2} and for two other parts, phase transition occurs at $J/2K \approx 3$. When the Ising perturbation is increased by $K$, at $K\approx J/6$ for the red and blue sublattices a phase transition occurs from topological order to ferromagnetic order but the green sublattice stays in topological phase. At $K\approx J/3$ topological order is completely removed and we have a ferromagnetic order. We can compare this result with that for the color code on the honeycomb lattice where phase transition occurs at $K \approx J/4.77$.
\subsection{The toric code on the honeycomb lattice}
\begin{figure}[t]
\centering
\includegraphics[width=8cm,height=3.5cm,angle=0]{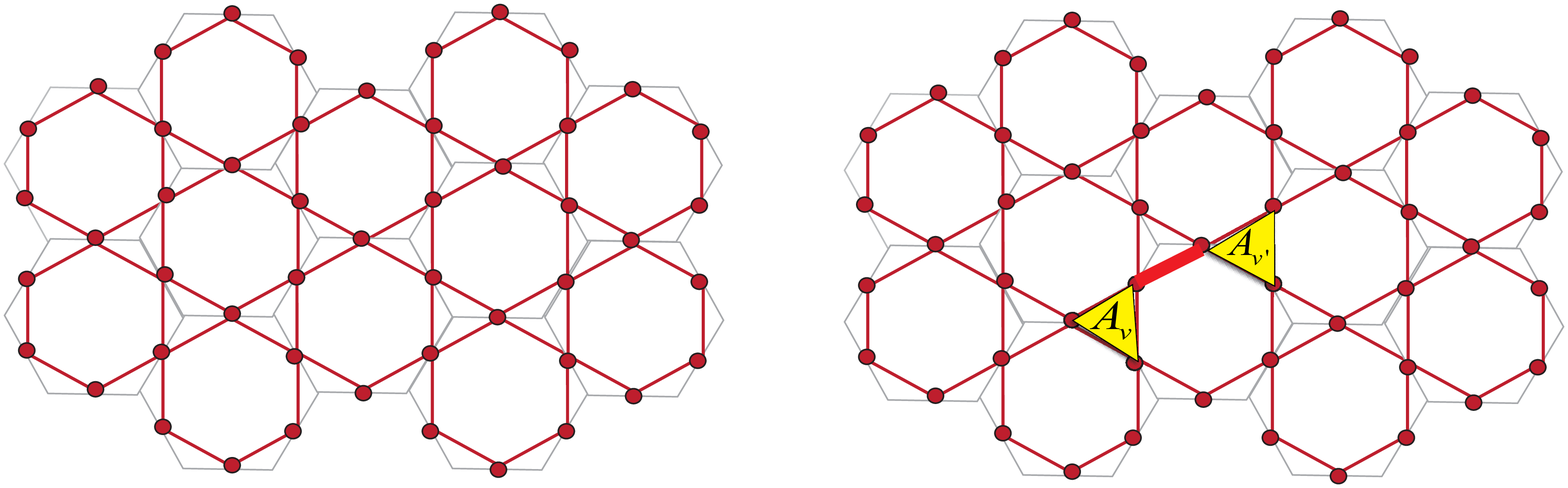}
\caption{(Color online) Left: Ising interaction pattern has been shown by a triangle-hexagonal lattice. Right: We denote two vertex operators by two dark triangles. An Ising interaction $Z_i Z_j $ between two real spins does not commute with two vertex operators on two it's endpoints and it is equivalent to an Ising interaction $X_v X_{v'}$ between corresponding virtual spins which are denoted by red circles.
} \label{tri}
\end{figure}

\begin{figure}[t]
\centering
\includegraphics[width=4cm,height=3.5cm,angle=0]{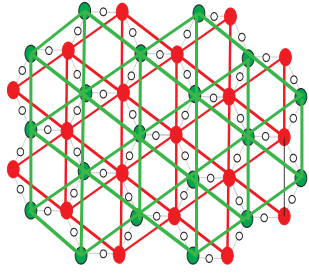}
\caption{(Color online) The virtual spins on vertices of lattice are denoted by two green and red colors. Ising pattern between real spins is converted to two independent Ising model on virtual spins.
} \label{triangle}
\end{figure}

In this subsection, we study the toric code on a honeycomb lattice with the Ising perturbation. To this end, we consider the nearest neighbors of each spin and we assume an Ising interaction between them. As it has been shown in Figure (\ref{tri}, left), if we illustrate each Ising interaction by a link, Ising interaction pattern is described by a triangle-hexagonal lattice which the spins live on it's vertices. In this new lattice the vertex operators of the toric code correspond to the triangular plaquettes and the plaquette operators of the toric code correspond to the hexagonal plaquettes. Finally the toric code with the Ising perturbation can be described by a Hamiltonian in the form of:
\begin{equation}\label{asd}
H=-J(\sum_{v}A_v +\sum_{p}B_p)-k\sum_{\la i,j\ra}Z_i Z_j .
\end{equation}
By the fact that the operator $Z_i Z_j $ commute with the operators $B_p$ and the non-trivial loop operators $L^{\sigma,Z}$
, we find that the ground state of the above Hamiltonian is in a subspace which is stabilized by $B_p$'s and $L^{\sigma,Z}$. Therefore, we select the following basis to re-write the Hamiltonian (\ref{asd}):
\begin{equation}\label{q1}
|\phi_{r_1 , r_2 , ...,r_N}\ra =\prod_{v}(1+(-1)^{r_v}A_v )|00...0\rangle .
\end{equation}
Then, we find new form of the Hamiltonian (\ref{asd}) in this basis. First, By the fact that $B_p$'s commute with all operators $A_v$, we find that $\sum_{p}B_p |\phi_{r_1 , r_2 , ...,r_N}\ra=N|\phi_{r_1 , r_2 , ...,r_N}\ra$. Second, we apply the operators $A_v$, by the fact that $A_v (1+(-1)^{r_{v}}A_v )=(-1)^{r_{v}}(1+(-1)^{r_{v}}A_v )$, we find that $\sum_{v}A_v |\phi_{r_1 , r_2 , ...,r_N}\ra =\sum_{v} (-1)^{r_v}) |\phi_{r_1 , r_2 , ...,r_N}\ra$. It is useful to describe the final relation in view of virtual spins. To this end, if we insert virtual spins on the vertices of the initial honeycomb lattice, the state $|\phi_{r_1 , r_2 , ...,r_N}\ra$ can treated as product state $| r_1 , r_2 , ...,r_N\ra$ on the virtual spins. Therefore, effect of the operator $A_v$ on the new basis is equivalent to effect of operator $Z_v$ on the virtual spin $r_v$.\\
Finally, we apply the Ising operator $Z_i Z_j$ to the basis (\ref{q1}). To this end, we consider a special link on the triangle-hexagonal lattice. Since the corresponding Ising operator to such a link does not commute with two vertex operators in two it's endpoints, see Figure (\ref{tri}, right), we find the following relation between each Ising operator and a vertex operator in one of the it's endpoints:
\begin{equation}
Z_i Z_j (1+(-1)^{r_v}A_v )= (1+(-1)^{r_v +1}A_v )Z_i Z_j .
\end{equation}
Therefore, effect of the operator $Z_i Z_j$ on the basis (\ref{q1}) leads to flip the virtual spins which live on two endpoints of the corresponding link. It means that the Ising interaction $Z_i Z_j$ on the real spins in the initial model is equivalent to an Ising interaction $X_v X_{v'}$ on the virtual spins. \\In the following, we should find effect of all Ising interactions in the initial model. To this end, according to Figure (\ref{triangle}), we color vertices of the initial honeycomb lattice by two different colors, blue and green. By attention to the above argument, each Ising interaction between the real spins in the initial lattice, similar to what has been shown in Figure (\ref{tri}, right), is equivalent to an Ising interaction $X_v X_{v'}$ between two virtual spins in the same color. Finally, the Ising interactions in the initial honeycomb lattice convert to the Ising interactions on two triangular lattices in two different colors and the Hamiltonian (\ref{asd}) in term of the virtual spins are wrote as follows:
\begin{equation}
H=-JN -(J\sum_{v \in T_B}Z_v +k\sum_{\la v,v' \ra \in T_B}X_v X_{v'})-(J\sum_{v \in T_G}Z_v +k\sum_{\la v,v' \ra \in T_G}X_v X_{v'}),
\end{equation}
where $T_B$ ($T_G$) refers to the blue (green) triangular lattice. Therefore, the initial Hamiltonian is converted to two independent Ising models in the transverse field on triangular lattices, see Figure (\ref{triangle}). As we pointed in the previous subsection, the phase transition point of this model occurs at ratio $J/k \approx  4.77 $. We can compare this result with the toric code on square lattice where the phase transition point is at $J/k \approx  6 $ \cite{karimipour}.\\
\subsection{The toric code on the triangular lattice}
\begin{figure}[t]
\centering
\includegraphics[width=4cm,height=3.5cm,angle=0]{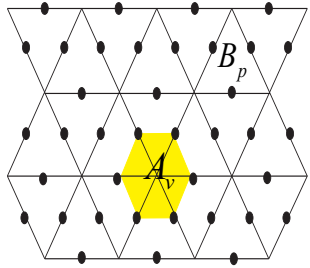}
\caption{(Color online) Plaquette and vertex operators has been shown.
} \label{tr}
\end{figure}

\begin{figure}[t]
\centering
\includegraphics[width=8cm,height=3.5cm,angle=0]{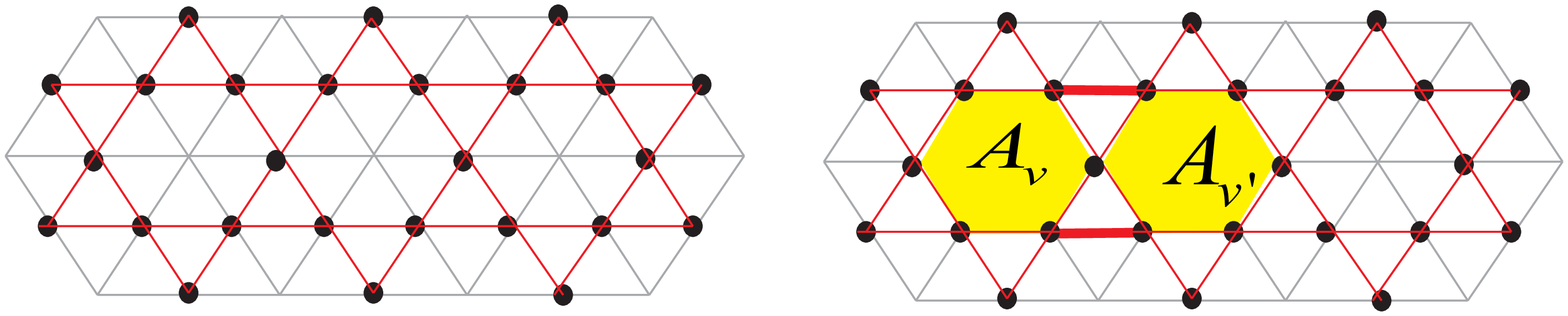}
\caption{(Color online) Left: Ising interaction pattern has been shown by a triangle-hexagonal lattice. Right: Corresponding Ising perturbation to a special link does not commute with two vertex operators in two endpoints of that.
} \label{tr2}
\end{figure}

\begin{figure}[t]
\centering
\includegraphics[width=4cm,height=3.5cm,angle=0]{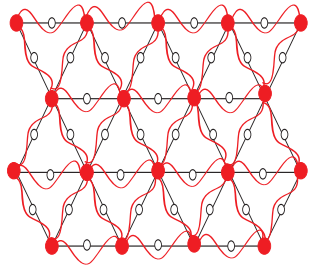}
\caption{(Color online) The toric code model in Ising perturbation is mapped to a Ising model in transverse field on virtual spins of a triangular lattice.
} \label{tr3}
\end{figure}

As a final case, we consider the toric code on the triangular lattice. In such a model, spins live on the edges of a triangular lattice and two kinds stabilizer operators are defined corresponding to plaquettes and vertices of the lattice, see Figure(\ref{tr}). Ising perturbation is applied between neighboring spins where pattern of the Ising interactions has been shown in Figure(\ref{tr2}, left). Similar to the previous subsection, we consider a non-local basis as:
\begin{equation}
|\phi_{r_1 , r_2 , ...,r_N}\ra =\prod_{v}(1+(-1)^{r_v}A_v )|00...0\rangle .
\end{equation}
In such a basis, the plaqutte operators $B_p$ are equal to Identity and the vertex operators $A_v$ are equal to Pauli operator $Z_v$ on virtual spins which live on vertices of the lattices.\\
For Ising interactions in the initial Hamiltonian, as it has been shown in Figure(\ref{tr2}, right), by the fact that each Ising operator $Z_i Z_j $ does not commute with two neighboring vertex operators $A_v$ and $A_{v'}$, we find that the Ising operator in the initial model is equal to an Ising operator $X_v X_{v'}$ in the new basis. If we apply such a transformation for all Ising operators, see Figure(\ref{tr3}), we will have a triangular lattice on virtual spins with Ising interaction $X_v X_{v'}$ between neighbor spins. Also there is an important point that two Ising interactions in the initial model are mapped to an new Ising interaction on virtual spins. Finally, the initial model in the new basis would be an Ising model in a transverse field as:
\begin{equation}
H=-kN-J\sum_{v}Z_v -2k \sum_{\la v, v' \ra} X_v X_{v'}.
\end{equation}
The phase transition point for this model is at $J/2k \approx  4.77$.
\begin{table}[t]
\centering
\begin{tabular}{ |p{2cm}||p{5cm}|p{5cm}|p{1.5cm}|  }
 \hline
 \multicolumn{2}{|c|}{$The~ topological~ code~in ~Ising ~perturbation$} & $Ising ~model~ in~ transverse~ field $   & $K/J$ \\
 \hline\hline
 \multirow{2}{4em}{$The~ color~code$} & $on~ honeycomb~ lattice$ &$on ~triangular~ lattice$& $\approx ~ 0.209$\\
  &$on~ square-octagonal~ lattice$ &$on~ square~ lattice$& $\approx ~ 0.333$\\
 \hline
 \multirow{3}{4em}{$The~ toric~code$}& $on~ square~ lattice$ &$on~ square~ lattice$& $\approx ~ 0.166$\\
&$on~ honeycomb~ lattice$ &$on ~triangular~ lattice$& $\approx ~ 0.209$\\
& $on~ triangular ~lattice$ &$on~ triangular~ lattice$&$ \approx ~ 0.104$\\
 \hline
\end{tabular}
\caption{The corresponding Ising model to each Topological codes and the phase transition points have been shown.}
\label{tab}
\end{table}

\section{Comments on results of the mapping}\label{sec3}

A summary of results of the previous section has been shown in Table (\ref{tab}). According to our derivations, two well known sets of topological codes on the most important lattices has been mapped to Ising model in transverse field on different lattices. \\
First, we compare the robustness of the TCI's on different lattices. The TCI on square lattice is mapped to IT on square lattice while the TCI on the honeycomb and the triangular lattices are mapped to IT on triangular lattice with different phase transition points. Therefore, the robustness of the TCI's depends on triangulation of the underlying lattice.\\
Second, we compare the CCI's on two different lattices. The CCI on honeycomb lattice is mapped to the IT on triangular lattice while the CCI on square-octagonal lattice is mapped to the IT on square lattice. Therefore, the robustness of the CCI's also depends on triangulation of the underlying lattice.\\
Third, we compare the robustness of the TCI's with the CCI's. The robustness of the TCI on square lattice is less than the robustness of the CCI on honeycomb lattice while the robustness is the same for the TCI and CCI on honeycomb lattice. Therefore, triangulation of the underlying lattice is important in comparison of two different topological codes so that there is not any way to compare the robustness of two different topological codes independent of their underlying lattices.\\
\section{Discussion}
In this paper, we derived the points of phase transition for two important topological codes on various lattices with the Ising perturbation, the toric code and the color code. To this end, we presented a detailed mapping between such models and the Ising model in transverse field. Our results explicitly showed that points of the phase transition and also robustness of perturbed topological codes depend on triangulation of lattices. Also, by comparing the robustness of the toric code and the color code, we showed that there is no way to compare the robustness of two different topological codes independent of triangulation of the underlying lattices.
\section*{Acknowledgements}
I would like to thank A. Poostforush and A. Montakhab for their help in finalization of the paper. Also I thank V. Karimipour for his good comments on the paper.

\end{document}